\documentclass[conference]{IEEEtran}
 \usepackage[cmex10]{amsmath}
\selectfont

\usepackage{amsmath}
\date{\ }

\newtheorem{theorem}{Theorem}
\newtheorem{lemma}{Lemma}

\begin{document}

\title{The best possible upper bound on the probability of undetected error for linear codes of full support}

\author{Torleiv Kl{\o}ve and Jinquan Luo, Department of Informatics, University of Bergen, N-5020 Bergen, Norway}

\maketitle

\begin{abstract}
There is a known best possible upper bound on the probability of undetected error for linear codes. 
The $[n,k;q]$ codes with probability of undetected error meeting the bound
have support of size $k$ only. In this note, linear codes of full support ($=n$) are studied. 
A best possible upper bound on the probability of undetected error for such codes is given, 
and the codes with probability of undetected error meeting 
this bound are characterized.
\end{abstract}

\section*{Upper bounds on $P_{\rm ue}(C,p)$ for linear codes $C$}

Let $n\ge k \ge 1$. An $[n,k;q]$ code is a linear code of length $n$ and dimension $k$ over the field $F_q$ of $q$ elements.

For an $[n,k;q]$ code $C$, the probability of undetected error $P_{\rm ue}(C,p)$ is the probability that a
codeword is changed to another codeword when transmitted over the $q$-ary symmetric channel. It is known, 
see \cite[Theorem 2.51]{K}, that

\begin{theorem}
If $C$ is an $[n,k;q]$ code, then
\begin{equation}
\label{gb}
P_{\rm ue}(C,p) \le (1-p)^{n-k}-(1-p)^n
\end{equation}
for all $p\in [0,(q-1)/q]$. Moreover, the bound is best possible since the bound is met with equality for all $p$
for the code $C_{n,k}$ generated by $[I_k|0_{k\times (n-k)}]$.
Here $I_k$ is the $k\times k$ identity matrix, and $0_{k\times (n-k)}$  is the $k\times (n-k)$ matrix with all entries zero.
\end{theorem}

It is known (see e.g. \cite{K} Theorem 2.1) that 
\[P_{\rm ue}(C,p)=(1-p)^n \left\{A_C\left( \frac{p}{(q-1)(1-p)}\right)-1  \right\}\]
where $A_C(z)$ is the weight distribution function of $C$.
In terms of the weight distribution, (\ref{gb}) is equivalent to 
\[A_C(z)\le A_{C_{n,k}}(z) \mbox{ for all }z\in [0,1].\]

For a code $C$ of length $n$, the support $\chi(C)$ is the set of positions $i$ such that $c_i\ne 0$ for
some codeword $(c_1,c_2,\ldots ,c_n)\in C$. The code has \emph{full support} if $|\chi(C)|=n$, that is,
for any position there is a codeword that is non-zero in this position. For example, the code 
$C_{n,k}$ has support $k$. 

In practical applications, one usually uses codes with full support. We expect to find a sharper upper bound on 
$P_{\rm ue}(C,p)$ for codes of full support. In this paper we find the following best possible 
upper bound on $P_{\rm ue}(C,p)$ for linear codes of full support.

\begin{theorem}
\label{th2} 
If $C$ is an $[n,k;q]$ code of full support, then
\[P_{\rm ue}(C,p) \le (1-p)^{n-k+1}+(q-1)^{k-n}p^{n-k+1}-(1-p)^n\]
for all $p\in [0,(q-1)/q]$. Moreover, the bound is best possible since the bound is met with equality for all $p$
for the code $D_{n,k,{\mathbf{v}}}$ 
generated by 
\[ \left[I_k\Bigl \lvert \begin{matrix} {\mathbf{v}} \\ 0_{(k-1)\times (n-k)} \end{matrix} \right],\] 
where ${\mathbf{v}}\in F_q^{n-k} $ is a vector of full support (that is, without zero in any position).
 Moreover, any code of full support meeting the bound is equivalent to $D_{n,k, {\mathbf{v}} }$ 
for some ${\mathbf{v}}$ of full support.
\end{theorem}

This bound is tighter than the bound (\ref{gb}). The improvement for $p\in (0,(q-1)/q)$ is
\[ p(1-p)^{n-k} \left \{1-\left( \frac{p}{(q-1)(1-p)}\right)^{n-k}\right\}.    \]

\section*{Proof of Theorem \ref{th2}}

The weight distribution of $D_{n,k, {\mathbf{v}} }$ is 
\[A_{D_{n,k,{\mathbf{v}} }}(z)= (1+(q-1)z)^{k-1}(1+(q-1)z^{n-k+1}) .\]

Therefore, Theorem \ref{th2} is equivalent to 
\begin{theorem}
\label{th3} 
If $C$ is an $[n,k;q]$ code of full support, then
\[A_C(z) \le (1+(q-1)z)^{k-1}(1+(q-1)z^{n-k-1}) \]
for all $z\in [0,1]$, with equality if and only if $C$ is equivalent to $ D_{n,k,{\mathbf{v}}} $ 
for some vector ${\mathbf{v}}$ of  full support. 
\end{theorem}

\begin{lemma}
\label{full}
An $[n,k;q]$ code $C$ has full support if and only if $C^\perp$ is an $[n,k,2;q]$ code, that is, it has minimum distance at least 2.
\end{lemma}
\begin{IEEEproof}
The result follows from the observation that if $i$ is not in the support, then the unit vector ${\mathbf{e}}_i$ is 
contained in $C^\perp$ and vice versa.
\end{IEEEproof}

By the MacWilliams theorem, if $C$ is an $[n,k;q]$ code, then
\begin{equation}
\label{mw}
A_{C^\perp}(z)=\frac{1}{q^k} \left(1+(q-1)z \right)^n A_C\left(\frac{1-z}{1+(q-1)z} \right).
\end{equation}

This implies that $A_{C_1}(z)\le A_{C_2}(z)$ for all $z\in [0,1]$ if and only if
$A_{C^\perp_1}(z)\le A_{C^\perp_2}(z)$ for all $z\in [0,1]$. 

Let $E_{n,k,{\mathbf{v}}}=D_{n,n-k,{\mathbf{v}}}^\perp$.  This code is generated by the matrix $[I_{k}|{\mathbf{v}}^t|0_{k\times (n-k-1)}]$.
 
Using (\ref{mw}), we see that
\begin{equation}
\label{ew}
A_{E_{n,n-k,{\mathbf{v}}} }(z)= \frac{1}{q} \Bigl\{ \bigl(1+(q-1)z\bigr)^{n-k+1}+(q-1)(1-z)^{n-k+1}\Bigr\}.
\end{equation}

Combining all these facts, we see that Theorem \ref{th3} is equivalent to the following (where we substitute $n-k$ for $k$).
\bigskip

\begin{theorem}
\label{th4} 
If $C$ is an $[n,k,2;q]$ code, then
\begin{equation}
\label{gb4}
A_C(z) \le f(z), 
\end{equation}
where \[f(z)=\frac{1}{q} \left\{ \left(1+(q-1)z\right)^{k+1}+(q-1)(1-z)^{k+1}\right\},\]
for all $z\in [0,1]$, with equality if and only if $C$ is equivalent to $E_{n,k,{\mathbf{v}}}$
for some vector ${\mathbf{v}}\in F_q^k$ of full support. 
\end{theorem}

Before proving this theorem, we give a couple of simple lemmas. For $z\in [0,1]$ we clearly have $z^i\ge z^j$ for $i\le j$.
This implies the following lemma.

\begin{lemma} 
\label{max un err} 
For $[n,k;q]$ codes $C$ and $C'$, if
\[\sum\limits_{i=1}^jA_i(C)\leq \sum\limits_{i=1}^jA_i(C')\]
for any $1\leq j\leq n$, then for all $z\in [0,1]$, we
have 
\[A_C(z)\le A_{C'}(z).\]
Moreover, we have equality for any $z\in (0,1)$ if and only if $A_i(C)=A_i(C')$
for all $i$, $1\leq i\leq n$.
\end{lemma}

\begin{lemma} 
\label{ewl}
Let ${\mathbf{v}}$ be a vector of full support. Then

\noindent {\rm a)}
\[ A_i(E_{n,k, {\mathbf{v}} }) = \frac{1}{q}\binom{k+1}{i} \left\{(q-1)^i+(q-1)(-1)^i  \right\}.\]
\noindent {\rm b)}
\begin{align}
\sum_{i=2}^{j} A_i(E_{n,k,{\mathbf{v}}}) &= \sum_{i=1}^{j-1}\binom{k}{i}(q-1)^i \nonumber \\
                        & +\frac{1}{q} \binom{k}{j} \left\{(q-1)^{j} +(-1)^j(q-1)\right\}. \label{sume}
\end{align}
\end{lemma}
\begin{IEEEproof}
We see that a) follows immediately from (\ref{ew}). From a) we get
\begin{align*}
\sum_{i=2}^{j} A_i(E_{n,k,{\mathbf{v}}}) =\, & \frac{1}{q} \sum_{i=2}^{j}\binom{k+1}{i}(q-1)^i\\
                                             & +  \frac{q-1}{q} \sum_{i=2}^{j}\binom{k+1}{i}(-1)^i. 
\end{align*}
Let
\[F(z)=\sum_{i=2}^{j}\binom{k+1}{i}z^i.\]
Then
\begin{align*} 
F(z) =\, & \sum_{i=2}^{j}\binom{k}{i}z^i + \sum_{i=2}^{j}\binom{k}{i-1}z^i  \\
=\, & \sum_{i=2}^{j}\binom{k}{i}z^i + \sum_{i=1}^{j-1}\binom{k}{i}z^{i+1}  \\
=\, & (z+1)\sum_{i=1}^{j-1}\binom{k}{i}z^i + \binom{k}{j}z^{j}-zk.  \\
\end{align*}
Hence
\begin{eqnarray*}
\lefteqn{q \sum_{i=2}^{j} A_i(E_{n,k,{\mathbf{v}}})} \\
    &=& F(q-1)+(q-1)F(-1) \\ 
    &=& q \sum_{i=1}^{j-1}\binom{k}{i}(q-1)^i + \binom{k}{j}(q-1)^{j}-(q-1)k  \\
    && + (q-1) \binom{k}{j}(-1)^j+(q-1)k.
\end{eqnarray*}
Hence, b) follows.
\end{IEEEproof}

We now give the proof of Theorem \ref{th4}.
\begin{IEEEproof}
Suppose $C$ is generated by $G=[I_k|Q]$ where
the rows of $Q$ are 
${\mathbf{v}}_1, {\mathbf{v}}_2,\cdots,
{\mathbf{v}}_k$ (and where ${\mathbf{v}}_i\neq
{\mathbf{0}}$ for $1\leq i\leq k$). Then for any ${\mathbf{x}}\in F_q^k$, the codeword 
$ {\mathbf{x}}G =( {\mathbf{x}} | {\mathbf{x}}Q) $ has
weight 
\[w({\mathbf{x}}G ) = w({\mathbf{x}} )+w({\mathbf{x}}Q).\]
Hence
\begin{equation}
\label{weight less j} 
\sum_{i=2}^j A_i(C) = S_1+S_2,  
\end{equation}

where
\begin{align*}
S_1 &= |\{{\mathbf{x}}\mid {\mathbf{x}}\neq
{\mathbf{0}},w({\mathbf{x}})\leq j-1, w({\mathbf{x}}Q)+w({\mathbf{x}})\leq j\}| \\
    &\le |\{{\mathbf{x}}\mid {\mathbf{x}}\neq
{\mathbf{0}},w({\mathbf{x}})\leq j-1\}| \\
    &=  \sum\limits_{i=1}^{j-1}\binom{k}{i}(q-1)^i,
\end{align*}
and
\[S_2= |\{{\mathbf{x}}\mid w({\mathbf{x}})= j,{ \mathbf{x}}Q={\mathbf{0}}.\}| \]
To evaluate $S_2$, we first choose
$j$ positions out of $k$, the number of choices is $\binom{k}{j}$.
Without loss of generality we can assume that 
${\mathbf{x}}=(x_1,x_2,\cdots, x_k)$, where  
$x_1, x_2, \cdots, x_j$ are
nonzero and $x_{j+1}=\cdots =x_k=0$. Then we have
\begin{equation}\label{reduce}
  \left\{ \begin{array}{ll}
&x_1, x_2,\cdots, x_j\neq 0\\
&x_1 {\mathbf{v}}_1+x_2{\mathbf{v}}_2+\cdots+x_j {\mathbf{v}}_j={\mathbf{0}}.
\end{array}
\right.
\end{equation}
Let $r$ be the rank of the matrix with rows 
 ${\mathbf{v}}_1, {\mathbf{v}}_2,\cdots, {\mathbf{v}}_j$.

If $r=1$, then for $1\leq i\leq j$, ${\mathbf{v}}_i=t_i {\mathbf{v}}_j$ for some
$t_i\in F_q^*$. Denote by $n_j$ the number of solutions of
$(\ref{reduce})$. For arbitrary nonzero elements $x_1, x_2,\cdots,
x_{j-1}$,
\begin{itemize}
  \item if $x_1 t_1+x_2 t_2+\cdots+x_{j-1} t_{j-1}=0$, then $(x_1, x_2, \cdots,
  x_{j-1})$ contributes $1$ to $n_{j-1}$.
  \item if $x_1 t_1+x_2 t_2+\cdots+x_{j-1} t_{j-1}\neq 0$, then 
  \[x_j=-x_1 t_1-x_2 t_2-\cdots-x_{j-1} t_{j-1}\]
   and $(x_1, x_2, \cdots, x_{j-1}, x_j)$ contributes $1$ to $n_{j}$.
\end{itemize}
Therefore we have $n_{j-1}+n_{j}=(q-1)^{j-1}$. This recurrence
relation  and the first term $n_1=0$ imply that
\begin{equation}\label{num nj}
n_j= \frac{1}{q} \left((q-1)^{j}+(-1)^j(q-1)\right).
\end{equation}

If $r\geq 2$, then we may assume that ${\mathbf{v}}_1$ and ${\mathbf{v}}_2$ are
linearly independent. For any fixed nonzero elements $x_3, \cdots,
x_j$, the equation 
\[x_1 {\mathbf{v}}_1+ x_2{\mathbf{v}}_2=-x_3
{\mathbf{v}}_3-\cdots-x_j{\mathbf{v}}_j\]
 has at most one solution. Therefore the
number of solutions of (\ref{reduce}) is at most $(q-1)^{j-2}$ which
is less than (\ref{num nj}) except when $q=2$, $j$ is odd, and
\[ {\mathbf{v}}_1+{\mathbf{v}}_2+\cdots+{\mathbf{v}}_j ={\mathbf{0}}.\]
 In this exceptional
case, $n_j=0<1=(q-1)^{j-2}$ and at least one of ${\mathbf{v}}_i$ has
Hamming weight at least $2$ (since an odd number of binary vectors of
weight $1$ can not have sum ${\mathbf{0}}$). We may assume
$w({\mathbf{v}}_j)\geq 2$. Choose ${\mathbf{x}}=(1,1,\cdots, 1,0)$. Then
$w({\mathbf{x}})=j-1$ and  
\[{\mathbf{x}}Q = {\mathbf{v}}_1+{\mathbf{v}}_2+\cdots+ {\mathbf{v}}_{j-1} =  {\mathbf{v}}_{j}.\] 
Hence
\[w({\mathbf{x}}G) = w({\mathbf{x}})+w({\mathbf{v}}_j) \ge j-1+2=j+1. \]
 Therefore, in the exceptional case, 
\[S_1< \sum_{i=1}^{j-1}\binom{k}{i}(q-1)^i.\]
In total,  by (\ref{weight less j}) we obtain
\begin{align}
\sum_{i=2}^j A_i(C) \leq & \sum_{i=1}^{j-1} \binom{k}{i}(q-1)^i \nonumber \\
                         & +\frac{1}{q} \binom{k}{j} \left((q-1)^{j} +(-1)^j(q-1)\right) \nonumber \\
                      =  & \sum_{i=2}^j A_i(E_{n,k,{\mathbf{v}}}) \label{weight j 2}
\end{align}
for $j\geq 2$ by (\ref{sume}).

 By Lemma
\ref{max un err} we get that $A_C(z)$ takes the maximal value
for any $z\in (0,1)$ if and only if $C$ is (equivalent to)
$E_{n,k,{\mathbf{v}}}$.
\end{IEEEproof}

\section*{On an older bound}

A special case of \cite[Theorem 2.51 ]{K} is equivalent to the statement that
\begin{equation}
\label{gam}
A_C(z)\le g(z) \stackrel{\rm def}{=} (1+(q-1)z)^k+k(q-1)(z^2-z)
\end{equation}
for all $[n,k,2;q]$ codes and all $z\in [0,1]$.
A simple proof goes as follows: we have
\[ w({\mathbf{x}}G) \ge w( {\mathbf{x}} ) \]
for all ${\mathbf{x}} \in F^k$. Moreover, if $w({\mathbf{x}})=1$,
then $w({\mathbf{x}}G) \ge 2$. Hence
\begin{align*}
A_C(z) \le &\, \sum_ {i=0}^k \binom{k}{i}((q-1)z)^i - k(q-1)z + k(q-1)z^2 \\
       = &\, (1+(q-1)z)^k +k(q-1)(z^2-z). 
\end{align*} 
Since (\ref{gb4}) is best possible for codes with minimum distance 2, it is clearly at least as good as (\ref{gam}).

If $k=0$, then $f(z)=g(z)=1$. If $k=1$, then \[f(z)=g(z)=1+(q-1)z^2.\]
If $k=q=2$, then $f(z)=g(z)=1+3z^2$. We will show that in all other cases, $g(z)>f(z)$.

\begin{theorem}
For $q\ge 2$ and $k\ge 1$  we have
\[g(z)-f(z)= \frac{q-1}{q}(1-z) \Bigl\{  \sum_{j=2}^k \binom{k}{j} \Bigl( (q-1)^j-(-1)^j\Bigr) z^j\Bigr\}.\]
In particular, $g(z)>f(z)$ for all $z\in (0,1)$, except when $q=k=2$ or $k=1$.
\end{theorem}
\medskip

\begin{IEEEproof}

\noindent $g(z)-f(z)$
\begin{align*}
  =\, & \Bigl(1+(q-1)z\Bigr)^k + k(q-1)(z^2-z) \\
      & -\frac{1}{q} \Bigl(1+(q-1)z\Bigr)^{k+1} - \frac{q-1}{q} (1-z)^{k+1}\\
  =\, & \frac{1}{q} \Bigl(1+(q-1)z \Bigr)^k \Bigl\{q-1-(q-1)z  \Bigr\} \\
      & -k(q-1)z(1-z) - \frac{q-1}{q} (1-z)^{k+1} \\
  =\, & \frac{q-1}{q}(1-z) \Bigl\{  \Bigl(1+(q-1)z \Bigr)^k - (1-z)^k-kqz\Bigr\} \\
  =\, & \frac{q-1}{q}(1-z) \Bigl\{  \sum_{j=0}^k \binom{k}{j} \Bigl( (q-1)^j -(-1)^j\Bigr) z^j -kqz \Bigr\} \\
  =\, & \frac{q-1}{q}(1-z) \Bigl\{  \sum_{j=2}^k \binom{k}{j} \Bigl( (q-1)^j -(-1)^j\Bigr) z^j\Bigr\}.
\end{align*}
In particular, if $q>2$, then $(q-1)^j-(-1)^j>0$ for all $j\ge 2$. If $q=2$, $(q-1)^j-(-1)^j>0$ if $j$ is odd. 
Hence, $g(z)>f(z)$, except when $k=q=2$ or $k=1$.
\end{IEEEproof}

\section*{Acknowledgement}

This work is supported by the Norwegian Research Council under the grant 191104/V30. 
The research of Jinquan Luo is also supported by NSF of China under grant 60903036, 
NSF of Jiangsu Province under grant 2009182 and the open research fund of National Mobile Communications Research Laboratory, 
Southeast University (No. 2010D12).


\end{document}